\documentstyle[12pt]{article} 
\begin{document} 
{ \pagestyle{empty} 
\vspace{10mm} 
\centerline{\Large \bf Comment on the Ghost State in 
the Lee Model}

\vspace{10mm}


\centerline{Kazuyasu Shigemoto \footnote{E-mail address:
shigemot@tezukayama-u.ac.jp}}
\centerline {{\it Department of Physics}}
\centerline {{\it Tezukayama University, Nara 631, Japan }}

\vspace{10mm}

\centerline{\bf Abstract} 

\vspace{5mm} 

We examine the ghost state in the Lee model, and give the physically 
meaningful interpretatation for norm of the ghost state.  According to 
this interpretation, the semi-positivity of the norm is guaranteed.

\vspace{2mm} 
\noindent
\hfil
\vfill

}

\vspace{10mm}

\noindent {\large \bf \S 1.\ \ Introduction} 

\vspace{2mm} 

\indent
The field theory is an essential tool to study particle physics,
because the particle theory only can treat the creation and 
the annihilation 
of particles.  While infinite quantities appear in the calculation
of the field theory, and the renormalization of physical quantities
becomes necessary.  But this renormalization scheme is 
an approximate and 
perturvative one. Then the Lee model\cite{Lee,Heisenberg} 
is introduced
to study the essense of the renormalization scheme in an exact form,
though this is a toy model in a sense that it restricts 
the interaction 
in a quite special form and also it is not relativistic invariant.

In this Lee model, there is difficulty that 
there appears the negative norm state, the ghost state, 
in the strong coupling region, which contradict with the 
physically meaningful condition. 

In this paper, we first review the Lee model and next 
give the physically
meanigful interpretation for the norm of the ghost state. 
According to this interpretation, the semi-positivity of the norm 
is guaranteed.  

\vspace{1cm}

\noindent {\large \bf \S 2.  Review of the Lee Model }

\vspace{2mm}

We breafly review the Lee model~\cite{Lee,Heisenberg}.
The Hamiltonian of the Lee model is given by 

\begin{eqnarray}
   &&H_0=m_{V_0} \int d \vec{p}  V^{\dag}(\vec{p}) V(\vec{p})
   +m_{N} \int d \vec{q}  N^{\dag}(\vec{q}) V(\vec{q})
   +\int d \vec{k}  \omega_k \theta^{\dag}(\vec{k}) \theta(\vec{k}), 
\label{II0} \\
   && H_I=\frac{g_0}{(2 \pi)^{3/2} } \int d \vec{k}  d \vec{p} 
  \frac{f(\omega_k)}{ \sqrt{2 \omega_k} } 
  \left\{V^{\dag}(\vec{p}) N(\vec{p}-\vec{k}) \theta(\vec{k})
        +N^{\dag}(\vec{p}-\vec{k}) V(\vec{p}) \theta^{\dag}(\vec{k})
  \right\},
\label{II1}
\end{eqnarray}

\noindent
where $\omega_k=\sqrt{\vec{k}^2+\mu^2}$, and $m_{V_0}$, 
$m_{N}$, $\mu$  represents bare masse of $V$, $N$, 
$\theta$ particles respectively.
The  vacuum state $|0>$ is defined by

\begin{eqnarray}
    V(\vec{p})|0>=N(\vec{q})|0>=\theta(\vec{k})|0>=0.
\label{II2}
\end{eqnarray}

One $N$ partcle state $N^{\dag}(\vec{q})|0> $ and 
one $\theta$ particle state $ \theta^{\dag}(\vec{k})|0> $
respectively is the trivial eigenstate of the total Hamiltonian

\begin{eqnarray}
    &&(H_0+H_I)N^{\dag}(\vec{q})|0>=m_{N}N^{\dag}(\vec{q})|0>, 
    \nonumber\\
    &&(H_0+H_I)\theta^{\dag}(\vec{k})|0>=\omega_k 
    \theta^{\dag}(\vec{k})|0>.
\label{II3}
\end{eqnarray}

While one $V$ particle state $ V^{\dag}(\vec{p})|0> $ itself is not
the eigenstate of the total Hamiltonian, but the linear 
combination of
 $V^{\dag}(\vec{p})|0>$ and $ \theta^{\dag}(\vec{k}) 
N^{\dag}(\vec{p}-\vec{k})|0> $
gives the eigenstate. 
We first examine what happens after applying the Hamiltonian 
to the one $V$ particle state $ V^{\dag}(\vec{p})|0>$  

\begin{eqnarray}
  &&H_0 V^{\dag}(\vec{p})|0>=m_{V_0} V^{\dag}(\vec{p})|0>, 
    \nonumber\\
  &&H_I V^{\dag}(\vec{p})|0>=\frac{g_0} {(2\pi)^{3/2}} 
   \int d \vec{k} \frac{f(\omega_k)}{\sqrt{2 \omega_k}} 
   \theta^{\dag}(\vec{k})
   N^{\dag}(\vec{p}-\vec{k})|0>. 
\label{II4}
\end{eqnarray}

We next examine what happens after applying the Hamiltonian 
to the state  
$ \theta^{\dag}(\vec{k}) N^{\dag}(\vec{p}-\vec{k})|0> $

\begin{eqnarray}
  &&H_0 \theta^{\dag}(\vec{k}) N^{\dag}(\vec{p}-\vec{k})|0>
  =(\omega_k+m_{N}) \theta^{\dag}(\vec{k}) 
   N^{\dag}(\vec{p}-\vec{k})|0>, 
  \nonumber\\
  &&H_I \theta^{\dag}(\vec{k}) N^{\dag}(\vec{p}-\vec{k})|0>
    =\frac{g_0} {(2\pi)^{3/2}} 
   \frac{f(\omega_k)}{\sqrt{2 \omega_k}} V^{\dag}(\vec{p})|0>.
\label{II6}
\end{eqnarray}

Then we try to find the renormalized V particle state 
$|V^{\rm ren}(\vec{p})>$, which is the eigenstate of 
the total Hamiltonian in the form 

\begin{eqnarray}
     |V^{\rm ren}(\vec{p})>=Z_{V}^{1/2} 
    \left\{ V^{\dag}(\vec{p})|0> +\int d \vec{k} \Phi(\vec{k}) 
    \theta^{\dag}(\vec{k}) N^{\dag}(\vec{p}-\vec{k})|0> \right\},
\label{II7}
\end{eqnarray}

\noindent and we denote the eigenvalue of the total Hamiltonian 
to be $m_{V}$ in the form 

\begin{eqnarray}
     (H_0+H_I)|V^{\rm ren}(\vec{p})>=m_{V} |V^{\rm ren}(\vec{p})>.
\label{II8}
\end{eqnarray}

Straightforward calculation gives 
 
\begin{eqnarray}
   &&(H_0+H_I)|V^{\rm ren}(\vec{p})>=Z_{V}^{1/2} 
    \left\{m_{V_{0}} V^{\dag}(\vec{p})|0> +
   \frac{g_0} {(2\pi)^{3/2}} 
   \int d \vec{k} \frac{f(\omega_k)}{\sqrt{2 \omega_k}} 
   \theta^{\dag}(\vec{k})
   N^{\dag}(\vec{p}-\vec{k})|0>  \right.    \nonumber \\
   && \left. +\int d \vec{k} \Phi(\vec{k}) \left( 
  (\omega_k+m_{N}) \theta^{\dag}(\vec{k}) 
   N^{\dag}(\vec{p}-\vec{k})|0> +\frac{g_0} {(2\pi)^{3/2}} 
   \frac{f(\omega_k)}{\sqrt{2 \omega_k}} 
    V^{\dag}(\vec{p})|0> \right)   \right\}                                         \nonumber \\
  &&=Z_{V}^{1/2} \left\{ \left( m_{V_{0}}+ 
  \frac{g_0} {(2\pi)^{3/2}} \int d \vec{k} \Phi(\vec{k})
  \frac{f(\omega_k)}{\sqrt{2 \omega_k}}\right) 
   V^{\dag}(\vec{p})|0> \right.     \nonumber \\
  &&+\left. \int d \vec{k}  \left( \frac{g_0} {(2\pi)^{3/2}} 
    \frac{f(\omega_k)}{\sqrt{2 \omega_k}}
    +(\omega_k+m_{N}) \Phi(\vec{k}) \right) 
   \theta^{\dag}(\vec{k}) N^{\dag}(\vec{p}-\vec{k})|0> \right\}.
\label{II81}
\end{eqnarray}

From Eq.(\ref{II8}), we obtain the relation

\begin{eqnarray}
  && m_{V}=m_{V_{0}}+ 
  \frac{g_0} {(2\pi)^{3/2}} \int d \vec{k} \Phi(\vec{k})
  \frac{f(\omega_k)}{\sqrt{2 \omega_k}} ,
\label{II82} \\
  &&m_{V}\Phi(\vec{k})=\frac{g_0} {(2\pi)^{3/2}} 
    \frac{f(\omega_k)}{\sqrt{2 \omega_k}}
    + (\omega_k+m_{N}) \Phi(\vec{k}).
\label{II83}
\end{eqnarray}

From Eq.(\ref{II83}), we have 

\begin{eqnarray}
  \Phi(\vec{k})=\frac{g_0}{(2 \pi)^{3/2}} 
  \frac{f(\omega_k)}{\sqrt{2 \omega_k}}
  \frac{1}{ m_V-m_N-\omega_k } .
\label{II9}
\end{eqnarray}

Substitution this $\Phi(\vec{k})$ into Eq.(\ref{II82}), we have 

\begin{eqnarray}
  m_V=m_{V_{0}}+\frac{{g_{0}}^2 }{(2 \pi)^3} \int d \vec{k} 
   \frac{f^2(\omega_k)}{2 \omega_k} \frac{1}{ m_V-m_N-\omega_k },
\label{II10}
\end{eqnarray}

\noindent which gives the mass renormalization relation 
\begin{eqnarray}
  \delta m_V=m_V -m_{V_{0}}=\frac{{g_{0}}^2 }{(2 \pi)^3} 
   \int d \vec{k} \frac{f^2(\omega_k)}{2 \omega_k} 
   \frac{1}{ m_V-m_N-\omega_k } .
\label{II11}
\end{eqnarray}

The norm of the renormalized vector state $|V^{\rm ren}(\vec{p})>$ 
gives the condition

\begin{eqnarray}
1=Z_V \left\{ 1 + \int d \vec{k} | \Phi(\vec{k}) |^2 \right\}.
\label{II12}
\end{eqnarray}

 Substituting the explicit form of $\Phi(\vec{k})$, we have 

\begin{eqnarray}
  Z_{V}^{-1}=1+\frac{{g_{0}}^2 }{(2 \pi)^3} \int d \vec{k} 
   \frac{f^2(\omega_k)}{2 \omega_k} 
   \frac{1}{ (m_V-m_N-\omega_k)^2 }.
\label{II13}
\end{eqnarray}

We connect the bare coupling $g_0$ and the renormalized coupling $g$
in the form $g^2=Z_V {g_{0}}^2$, which comes from the condition that
the $ N +\theta \rightarrow N+\theta$ scattering amplitude is 
expressed only with renormalized quantities. 

Using this relation, we have 

\begin{eqnarray}
  Z_{V}^{-1}=1+Z_{V}^{-1} \frac{ g^2 }{(2 \pi)^3} \int d \vec{k} 
   \frac{f^2(\omega_k)}{2 \omega_k} 
    \frac{1}{ (m_V-m_N-\omega_k)^2 },
\label{II14}
\end{eqnarray}

\noindent which can be rewritten in the form 

\begin{eqnarray}
  Z_{V}^{-1} \left( 1-\frac{ g^2 }{(2 \pi)^3} \int d \vec{k} 
   \frac{ f^2(\omega_k)}{2 \omega_k} 
   \frac{1}{ (m_V-m_N-\omega_k)^2 } 
   \right)=1.
\label{II15}
\end{eqnarray}

Thus we can solve $Z_{V}^{-1}$ in the form  
\begin{eqnarray}
  Z_{V}^{-1}= 1/ \left(1-\frac{g^2 }{(2 \pi)^3} \int d \vec{k} 
   \frac{f^2(\omega_k)}{2 \omega_k} 
   \frac{1}{ (m_V-m_N-\omega_k)^2 } \right),
\label{II16}
\end{eqnarray}

\noindent which gives 

\begin{eqnarray}
  Z_{V}=1-\frac{g^2 }{(2 \pi)^3} \int d \vec{k} 
   \frac{f^2(\omega_k)}{2 \omega_k} 
   \frac{1}{ (m_V-m_N-\omega_k)^2 }.
\label{II17}
\end{eqnarray}

\vspace{1cm}

\noindent {\large \bf \S 3.  Physical Interpretation for 
the Norm of the Ghost State}

\vspace{2mm}

Because $Z_V$ is the probability to exist bare 
$V$-particle state $|V(\vec{p})>$ in the renormalized $V$-particle
state $|V^{\rm ren}(\vec{p})>$, 

\begin{eqnarray}
 0 \le Z_V \le 1,
\label{III1}
\end{eqnarray}

\noindent must be satisfied.

In the strong coupling region, 

\begin{eqnarray}
 \frac{g^2 }{(2 \pi)^3} \int d \vec{k}  
   \frac{f^2(\omega_k)}{2 \omega_k} 
   \frac{1}{ (m_V-m_N-\omega_k)^2 } 
  >1, 
\label{III2}
\end{eqnarray}

\noindent $Z_{V}$ becomes negative by using the relation

\begin{eqnarray}
  Z_{V}=1-\frac{g^2 }{(2 \pi)^3} \int d \vec{k} 
   \frac{f^2(\omega_k)}{2 \omega_k} 
   \frac{1}{ (m_V-m_N-\omega_k)^2 },
\label{III3}
\end{eqnarray}

\noindent under the standard interpretation, but which contradict 
with the physically meaningful condition 
Eq.(\ref{III1}), which is the origin of the ghost problem in 
the Lee model. 

We can give another physically meaningful interpretation 
for that norm.

Let's denote 
\begin{eqnarray}
   x=\frac{g^2 }{(2 \pi)^3} \int d \vec{k}  
   \frac{f^2(\omega_k)}{2 \omega_k} 
   \frac{1}{ (m_V-m_N-\omega_k)^2 },
\label{III4}
\end{eqnarray}

and we interprete Eq.(\ref{II16}) as 

\begin{eqnarray}
  Z_{V}^{-1}= 1/(1-x)=1+x+x^2+x^3+x^4+\cdots \quad(x>1). 
\label{III5}
\end{eqnarray}

Then we have $Z_V^{-1}=1+x+x^2+x^3+x^4+\cdots$
$=+\infty$ $({\rm with}\ x>1)$, 
which means $Z_V=1/(1+x+x^2+x^3+x^4+\cdots)$ 
$=0$  $({\rm with}\ x>1)$ instead of taking negative value. 

We interprete that the expression $1/(1-x)$ 
$({\rm with}\ x>1)$ is the 
symbol to express infinity here, that is, we interprete 
$1/(1-x)=1+x+x^2+x^3+x^4+\cdots$ $({\rm with}\ x>1)$.  
We can make an intuitive interpretation why 
$1/(1-x)$  $({\rm with}\ x>1)$  can express infinity.
If we devide $1$ by $0=(1-1)$, we have infinity 
$1/(1-1)=1+1+1+ \cdots$. 
Then if we devide $1$ by $(1-x)$ $({\rm with}\ x>1)$ "the number 
with the magnitude being smaller than $0$", 
we obtain another infinity 
$1/(1-x)=1+x+x^2+x^3+x^4+\cdots$ $({\rm with}\ x>1)$,  
which is the larger infinity than the former infinity 
$1/(1-1)=1+1+1+ \cdots$. 
Of course, there is no actual number with 
"the magnitude being smaller than $0$". 
It is the number living in the world of ideal, that is, 
it appears only in connection with infinity.
\vspace{1cm}

\noindent {\large \bf \S 5. Summary }

\vspace{2mm}

We examine the ghost state in the Lee model, and give the 
physically meanigful 
interpretatation for norm of the ghost state. According to this 
interpretation, the semi-positivity of the norm is guaranteed. 
Depending on the situation,
$1/(1-x)$ $({\rm with}\ x>1)$ represent finite number or infinity 
of the form $1/(1-x)=1+x+x^2+x^3+x^4+\cdots$.  For the problem 
of the ghost state in the Lee model, we interprete 
$1/(1-x)$ $({\rm with}\ x>1)$ as the symbol 
to express infinity, and obtain the physically meaningful result.
It may be useful to classify many infinities by symbolically 
express these infinities in the form $1/(1-x)$ $({\rm with}\ x>1)$.

\vspace{10mm}


\end{document}